\documentclass[fleqn,twoside]{article}
\usepackage[english,danish]{babel}
\usepackage{espcrc2}
\usepackage{amsmath}
\usepackage{cite}
\usepackage{amssymb}
\usepackage{latexsym}
\usepackage{epic}
\usepackage{graphics}
\usepackage{epsfig}

\newcommand{\lrx}[1]{\left({#1}\right)}

\title{Numerical Results on the Non--Commutative $\lambda\phi^4$ Model 
\thanks{Talk presented by F. Hofheinz at Lattice 2003.
\newline \hspace*{0.9mm} Preprint HU-EP-03/66.}}

\author{W.\ Bietenholz \address{\vspace{-.25cm} Institut f\"ur Physik, Humboldt--Universit\"at zu Berlin, 
    Newtonstr. 15, D--12489 Berlin, Germany},
  F.\ Hofheinz $^{\mbox{\scriptsize a,}}$\address{\vspace{-.25cm} Fachbereich Physik, Freie Universit\"at Berlin, 
    Arnimallee 14, D--14195 Berlin, Germany}
  and J.\ Nishimura \address{ High Energy Accelerator Research Organization (KEK), 1-1 Oho, Tsukuba 305-0801, Japan}}

\begin{document}

\begin{abstract}
  The UV/IR mixing in the $\lambda\phi^4$ model on a non-commutative
  (NC) space leads to new predictions in perturbation theory,
  including Hartree--Fock type approximations.
  Among them there is a changed phase diagram and an unusual
  behavior of the correlation functions. In particular this mixing leads
  to a deformation of the dispersion relation. We present numerical
  results for these effects in $d=3$ with two NC coordinates.
  \vspace{-.3cm}
\end{abstract}

\maketitle

\section{INTRODUCTION}

Field theories defined on a NC geometry are highly fashionable, in particular
because they arise from a low energy limit of string theory
\cite{Seiberg:1999vs}.

A NC space may be defined by the non--commutativity of some
of its coordinates 
\begin{equation}  
  \label{NCspace}
  [\hat x_{\mu}, \hat x_{\nu} ] = i\Theta_{\mu\nu} \,\overset{\text{2d}}{=}\, i \theta \epsilon_{\mu \nu}\,.
\end{equation}
For a review of NC field theories, see Ref.\ \cite{Szabo:2001kg}.

\section{THE NC $\lambda\phi^4$ MODEL}
\label{sec:ph}

The extension of actions of commutative field theories to their NC
counterparts can be realized by replacing all products between fields
by the {\em star--product}
\begin{equation}
  f(x) \star g(x) = e^{\frac{1}{2} i \Theta_{\mu \nu}
    \frac{\partial}{\partial x_{\mu}} \frac{\partial}{\partial y_{\nu}}}f(x) g(y) \vert_{x=y}\,.
\end{equation}
In the NC $\lambda\phi^4$ model this replacement leads to the action
\begin{equation}
  \label{eq:phi-action}
  S\!=\!\!\int\! d^dx\!\left[\frac{1}{2}\partial_\mu \phi\,\partial_\mu 
    \phi\!+\!\frac{m^2}{2}\phi^2\!+\!\frac{\lambda}{4}\phi\star\phi\star\phi\star\phi\right]\,,\nonumber
\end{equation}
where only the interaction term requires the star--product.

In perturbation theory the one loop contribution to the 1~PI two point
function splits into two parts coming from the planar and the
non--planar graphs. The planar terms are proportional to their (UV
divergent) commutative counterparts \cite{Filk:1996dm}. In the case of
the non--planar graphs the momentum cut--off $\Lambda$ is replaced by
an ef\-fective cut--off $\Lambda_\text{eff}$ \cite{Minwalla:1999px},
\begin{equation}
  \label{eq:eff-cutoff}
  \Lambda_\text{eff}^2=\frac{1}{\frac{1}{\Lambda^2}+\theta^2p^2}\,,
\end{equation}
where $p$ denotes the incoming momentum.
The cut--off $\Lambda$ may be safely send to infinity, leading to
a UV finite contribution. However, the UV divergences reappear as IR
divergences in the limit $p\to0$. This mixing
of UV and IR effects still causes serious problems in a perturbative 
treatment of NC field theories beyond one loop.

We studied the mixing of divergences non--perturbatively in the 3d
model, with a commutative time direction. 
\footnote{For corresponding studies in $d=2$, see Refs.\ 
  \cite{Ambjorn:2002nj,Bietenholz:2002ev}.}
To avoid the (CPU) time consuming lattice version of the
star--product, we mapped the system on a dimensionally reduced model
\cite{Ambjorn:1999ts}. Here the scalar fields $\phi(\vec{x},t)$
defined on a $N^2\times T$ lattice are mapped on $N\times N$ Hermitian
matrices $\hat{\phi}(t)$. Their action reads
\begin{eqnarray}
  \label{eq:reduced-action}
  S[\hat{\phi}]&\hspace{-.5cm}=\vspace{-.1cm}N\textrm{Tr}\sum_{t=1}^{T} \biggl[
  \frac{1}{2}\sum_\mu\left(\Gamma_\mu\hat{\phi}(t)\,
    \Gamma_\mu^\dagger\!-\hat{\phi}(t)\right)^2\\
  &\hspace{-.5cm}+\frac{1}{2}\left(\hat{\phi}(t\!+\!1)-\hat{\phi}(t)\right)^2\!\!+
  \!\frac{m^2}{2}\hat{\phi}^2(t)+\frac{\lambda}{4}\hat{\phi}^4(t)
  \biggl]\nonumber\,,
\end{eqnarray}
where the twist--eaters $\Gamma_\mu$ are defined by
\begin{eqnarray}
  \label{eq:twisteaters}
  && \hspace*{-7mm}
  \Gamma_\mu\Gamma_\nu=Z^*_{\mu\nu}\Gamma_\nu\Gamma_\mu\,,\quad\text{where}\\
  &&\hspace*{-7mm}Z_{\mu \nu} = e^{\pi i\frac{N+1}{N} } = Z_{\nu \mu}^{*}
  \quad(\mu < \nu )\quad \text{is the twist.} \nonumber
\end{eqnarray}
This implies $\theta=\frac{1}{\pi}Na^2$, where $a$ is the lattice
spacing.  For this action we studied the phase diagram and the
dispersion relation.

\vspace{-.2cm}
\section{THE PHASE DIAGRAM}
\begin{figure}[htbp]
  \vspace{-.5cm}
  \centering
  \includegraphics[width=\linewidth]{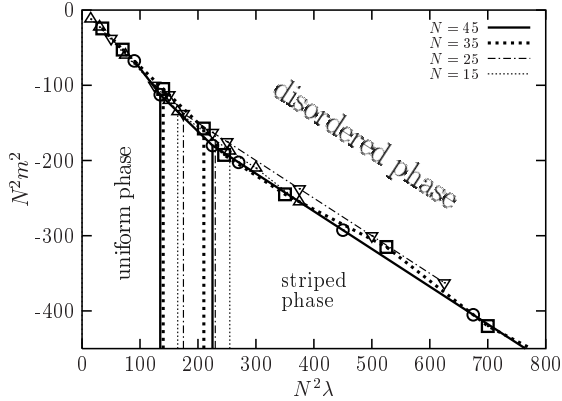}

  \vspace{-1cm}
  \caption{\it The phase diagram of the 3d NC $\lambda\phi^4$ model.
    The connected symbols show the separation line between disordered
    and ordered regime, and the vertical lines mark the transition
    region between uniformly ordered and striped phase.}
  \label{fig:phase}
  \vspace{-.6cm}
\end{figure}

Based on the momentum dependent order parame\-ter
$\langle M(k) \rangle$, with
\footnote{$\tilde{\phi}(\vec{p},t)$ is the spatial Fourier transform of $\phi(\vec{x},t)$.}
\begin{equation}
  \label{eq:orderparameter1}
  M(k):=\frac{1}{NT} \max_{|\vec{p}|=k}
  \bigl|\sum_t\tilde{\phi}\;(\vec{p},t)\bigl| \,,
\end{equation}
we studied the phase diagram in the $\lambda$--$m^2$ plane. 
Our results for various values
of $N=T$ are shown in Fig.\ \ref{fig:phase}.  
We identify a clear separation line (connected symbols) between the
disordered phase and the ordered regime. The ordered regime splits
into a uniformly ordered phase and a striped phase, where the
transition region is marked by two vertical lines for each value of
$N$.

\begin{figure}[htbp]
  \centering
  \includegraphics[width=.45\linewidth]{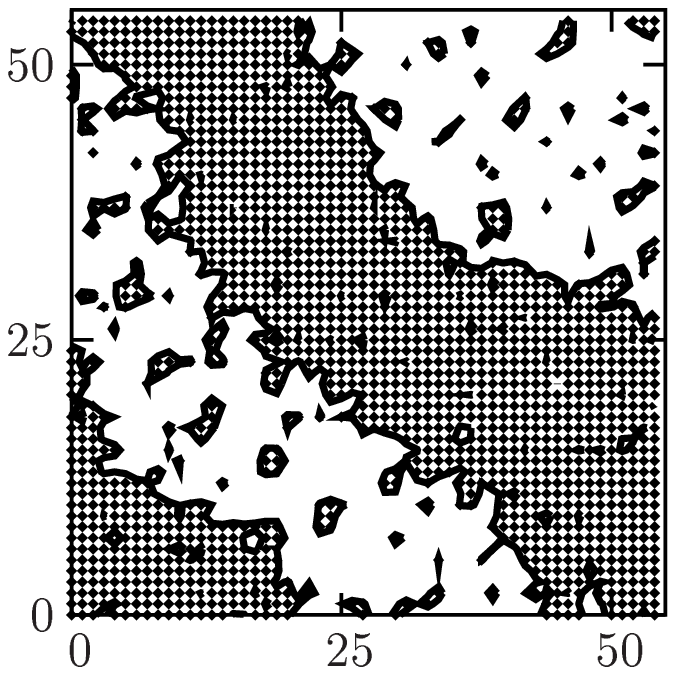}%
  \hspace{.5cm}\includegraphics[width=.45\linewidth]{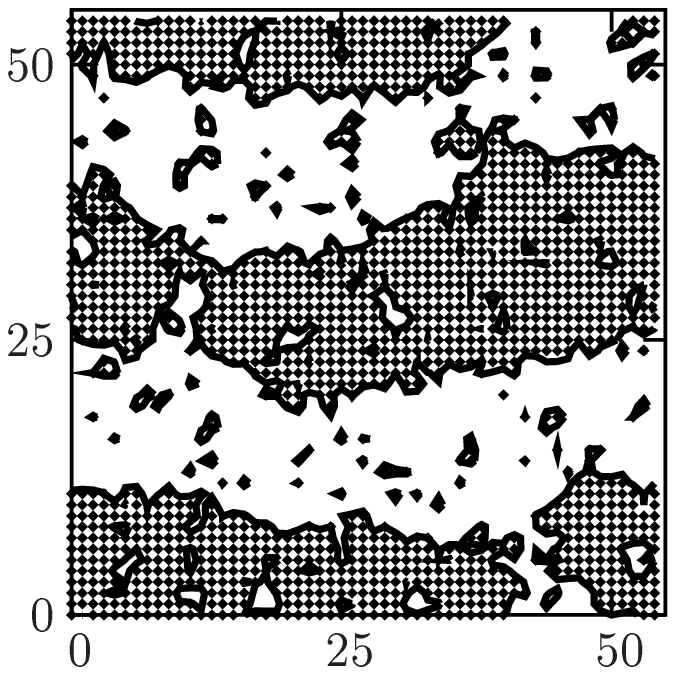}

  \vspace{-.9cm}
  \caption{\it Snapshots of single configurations $\phi(\vec{x},t)$
  at a certain time $t$, for $N=T=55$.}
  \label{fig:snapshot}
  \vspace{-.7cm}
\end{figure}

To illustrate the striped phase we present in Fig.\ \ref{fig:snapshot}
snapshots of single configurations, which represent the ground state
in this phase in the $x_1$--$x_2$ plane at fixed time $t$
for $N=55$ at $\lambda=50$ and $m^2=-15$. The dotted areas indicate
$\phi>0$ and in the blank areas $\phi$ is negative.  Here we show
configurations with two diagonal stripes resp.\ four stripes parallel to
the $x_1$ axis. At smaller values of the coupling $\lambda$ or smaller
system size $N$ we also find two stripes parallel to one of the axis
\cite{Bietenholz:2002ev}.

These results agree qualitatively with the conjecture by Gubser and
Sondhi, who predicted the occurrence of a striped phase
\cite{Gubser:2000cd}. To complete the agreement the striped phase has
to survive the continuum limit, where the number of stripes should
diverge, such that the width of the stripes remains finite.

\section{DISPERSION RELATION}
\vspace{-.1cm}
The star--product breaks explicitly the Lorentz symmetry, which leads
to a deformation of the standard dispersion relation. The one loop
result for this relation  reads \cite{Minwalla:1999px}
\begin{equation}
  \label{eq:deformation}
  E(\vec{p}\,)^2=\vec{p}^{\,2}+M_\text{eff}^2+\xi\lambda\Lambda_\text{eff}\,e^{-m/\Lambda_\text{eff}}\,,
\end{equation}
where $\Lambda_\text{eff}$ is defined in Eq.\ (\ref{eq:eff-cutoff}).
The deformation causes a shift in the energy minimum from
$\vec{p}=\vec{0}$ to non--zero momenta.

We investigated numerically the energy--momentum relation in the
disordered phase. The energy $E(\vec{p})$ can be computed from the
correlator
\begin{equation*}
  \label{eq:phi-gm}
  G(\vec{m},\tau)=\frac{1}{N^2T}\sum_t\left\langle\text{Re}\lrx{\tilde{\phi}^*(\vec{m},t)
      \tilde{\phi}(\vec{m},t+\tau)}\right\rangle,
\end{equation*}

\vspace{-.1cm}
\noindent where the physical momenta are given by $\vec{p}=2\pi\,\vec{m}/N$.
This correlator behaves like a {\tt cosh} 
\begin{equation}
  \label{eq:phi-gm-behavior}
  G(\vec{m},\tau) \propto \lrx{e^{-E(\vec{p})\,\tau}+e^{-E(\vec{p})(T-\tau)}}\,,
\end{equation}

\begin{figure}[htbp]
  \vspace{-.5cm}
  \centering
  \includegraphics[width=.95\linewidth]{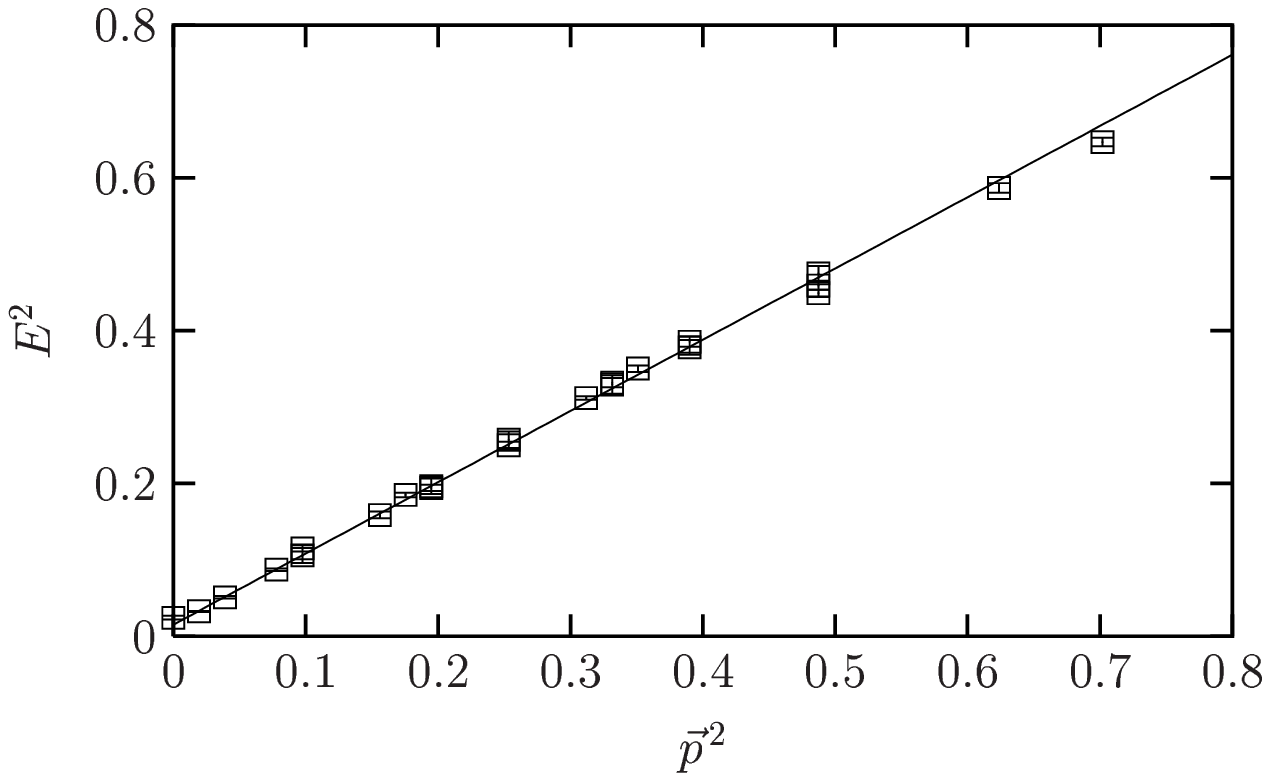}\\
  \includegraphics[width=\linewidth]{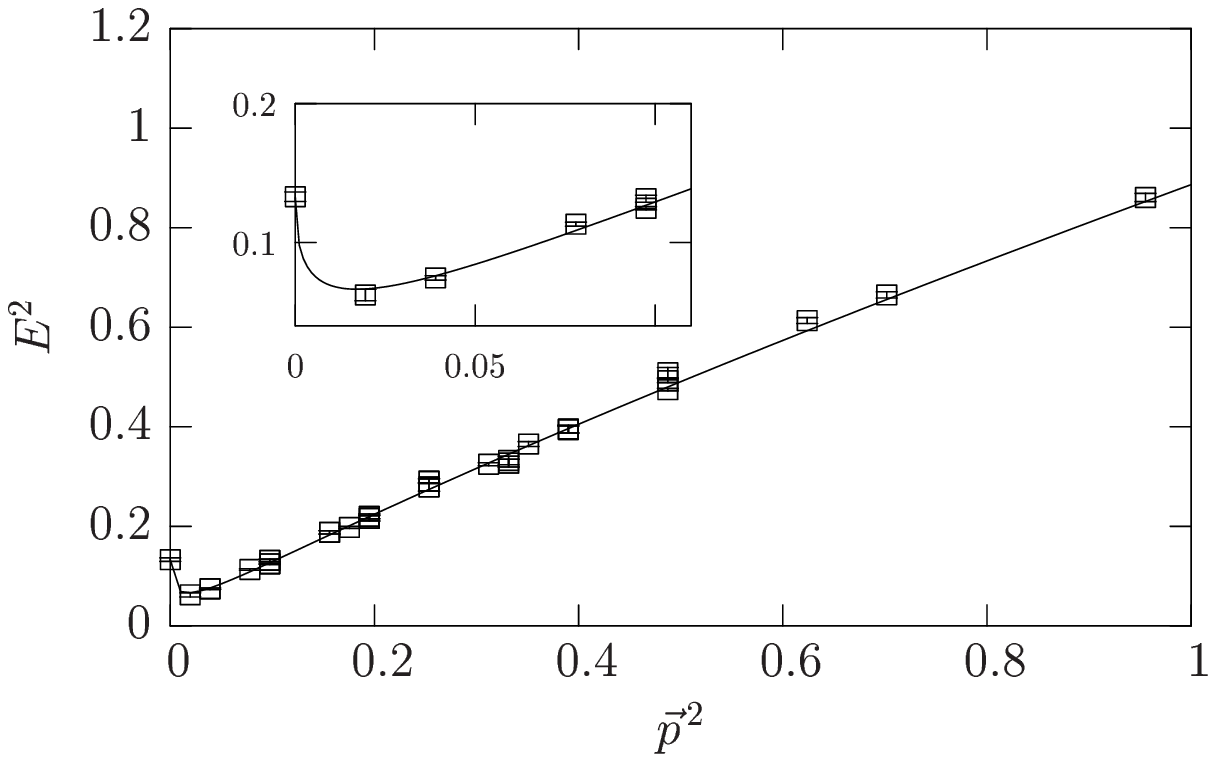}

  \vspace{-1cm}
  \caption{\it The dispersion relation in the disordered phase at $N=45$. Close to the 
    uniform phase at $N^2\lambda=90,\,N^2m^2=-22.5$ (on top) and close
    to the striped phase at $N^2\lambda=450,\,N^2m^2=-360$ (below).}
  \label{fig:disp1}
  \vspace{-.7cm}
\end{figure}

\vspace{-.1cm}
\noindent and the study of its decay allows to extract the energy. In Fig.\ 
\ref{fig:disp1} (on top) the system is close to the
uniformly ordered phase transition.
Here the square of the energy is linear in $\vec{p}^{\,2}$ as in a
Lorentz invariant theory. Close to the striped phase (Fig.\ 
\ref{fig:disp1} below) the situation is changed. We see a clear
deviation from Lorentz symmetry. The minimum of the energy is now at
the lowest non--zero (lattice) momentum and thus there will be two
stripes parallel to the axes in the non--uniform phase.

\begin{figure}[htbp]
  \vspace{-.7cm}
  \centering
  \includegraphics[width=\linewidth]{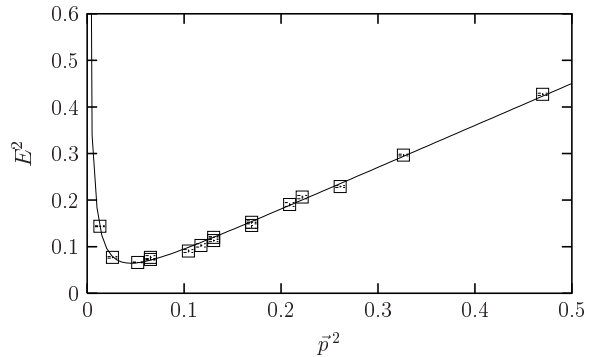}

  \vspace{-1cm}
  \caption{{\it The dispersion relation in the disordered phase 
      at $N=55$, $\lambda=50,\,m^2=-15$.}}
  \label{fig:disp2}
  \vspace{-.8cm}
\end{figure}

In Fig.\ \ref{fig:disp2} the results at very large coupling $\lambda$,
far outside the phase diagram in Fig.\ \ref{fig:phase}, are shown.
Now the energy minimum is shifted to larger momenta, leading to the
more complicated patterns in the striped phase as in Fig.\ 
\ref{fig:snapshot}.

In Figs.\ \ref{fig:disp1} and \ref{fig:disp2} the solid lines are fits
to the one loop result for the energy in Eq.\ (\ref{eq:deformation}).

\vspace{-.4cm}
\section{CONCLUSIONS}

\vspace{-.1cm}
We studied numerically the effects of UV/IR mixing in the 3d NC
$\lambda\phi^4$ model. For the phase diagram we found that the ordered
regime is split into an Ising type phase for small coupling $\lambda$
and a striped phase for larger coupling. The patterns in the striped
phase become more complex when $\lambda$ or the system size $N$ is
increased. These results are in qualitative agreement with the
conjecture of Gubser and Sondhi, if this type of stripes survives
the large $N$ limit.

The energy--momentum relation behaves as predicted from one loop
perturbation theory. This is a remarkable result, since due to the
UV/IR mixing there could be strong effects from higher order
calculations. Our results imply that such effects do not change the
results qualitatively. However, for final conclusions one has to
perform the continuum limit \cite{Bietenholz:xxx}.


\end{document}